\documentclass[12pt,preprint]{aastex}

\newcommand{\msun}{$M_{\odot}\ $}

\newcommand{\kms}{km s$^{-1}$~}

\shorttitle {Rapid rotation in the GC systems of Virgo dEs}
\shortauthors {Beasley et al.}

\begin{document}

\title {Evidence for the disky origin of luminous Virgo dwarf ellipticals 
from the kinematics of their globular cluster systems\altaffilmark{1}}

\author{Michael A. Beasley, A. Javier Cenarro}
\affil{Instituto de Astrof\'isica de Canarias,  V\'ia L\'actea s/n, 38200 La Laguna, 
Tenerife, Spain}
\email{beasley@iac.es, cenarro@iac.es}

\author{Jay Strader\altaffilmark{2}}
\affil{Harvard-Smithsonian Center for Astrophysics, 60 Garden St., Cambridge, MA 02138}
\email{jstrader@cfa.harvard.edu}

\and

\author{Jean P. Brodie}
\affil{UCO/Lick Observatory, University of California, Santa Cruz, CA 95064}
\email{brodie@ucolick.org}

\altaffiltext{1}{Some of the data presented herein were obtained at the
W.M. Keck Observatory, which is operated as a scientific partnership among 
the California Institute of Technology, the 
University of California and the National Aeronautics and Space Administration. 
The Observatory was made possible by the generous financial support of the W.M. 
Keck Foundation.}
\altaffiltext{2}{Hubble Fellow}

\begin{abstract}

We report evidence for dynamically significant rotation in the globular 
cluster systems of two luminous Virgo dwarf ellipticals, VCC1261 and VCC1528.
Including previous results for VCC1087, the 
globular cluster systems of all three Virgo dwarf ellipticals studied 
in detail to date exhibit $v_{\rm rot}/\sigma_{\rm los} >1$.
Taking the rotation seen in the globular clusters as a maximal disk rotation, 
and accounting for the possible fading of any hypothetical progenitor galaxy, 
we find all three dEs lie on the $r$-band Tully-Fisher relation.
We argue that these data support the hypothesis that luminous 
dEs are the remnants of transformed disk galaxies.
We also obtained deep, longslit data for the stars in VCC1261 and VCC1528. Both 
these galaxies show rapid rotation in their inner regions, with spatial 
scales of $\sim0.5$ kpc. These rotation velocities are surprisingly similar  
to those seen in the GC systems. 
Since our longslit data for Virgo dEs extend out to 1--2 effective radii  
(typical of deep observations), whereas the globular clusters extend out to 4--7 
effective radii, we conclude that non-detections of rotation in many luminous
dEs may simply be due to a lack of radial coverage in the stellar data, and 
that globular clusters represent singularly sensitive probes of the dynamics 
of dEs. Based on these data, we suggest that gas disks are significant 
sites of globular cluster formation in the early universe.
\end{abstract}

\keywords{galaxies: star clusters -- galaxies: dwarf -- galaxies: kinematics and dynamics}

\section {Introduction} \label{sec: 1}

Growing observational evidence suggests that a significant fraction
of dwarf elliptical galaxies (dEs) are the remnants of
transformed late-type disk galaxies (Kormendy 1985). 
Some of the most striking evidence for this are features such as vestigal disk 
and bar structures observed in many present-day dEs (e.g., Jerjen et al. 2000; 
Barazza et al. 2002). 
In a comprehensive analysis, Lisker et al. (2006) analyzed multi-band SDSS 
imaging of 476 Virgo cluster dEs. These authors found that more than 50\% of 
luminous Virgo dEs (M$_B<-16$) show evidence for disk features. 
These disky dEs (``dEdis'' to use the term of Lisker et al. 2006) also exhibit
different clustering properties to ``normal'' dEs; Virgo dEdis appear relatively
unclustered and follow the spatial distributions of irregulars and spiral
galaxies, whereas dEs are highly clustered like  E/S0 galaxies. Moreover, 
the luminosity functions of dEdis and dEs seem to differ, with 
the number fraction of dEdis relative to dEs declining sharply with decreasing 
luminosity.

Perhaps less conspicuous than overt spiral structure, but equally compelling evidence 
for a disky origin for some dEs has come from dE kinematics.
Integrated light studies suggest that a number of dEs 
are rotationally supported (Pedraz et al. 2002; Simien \& Prugniel 2002; 
Geha et al. 2003; van Zee et al. 2004)\footnote{The exact definition of 
rotational support varies among authors. Here, we consider
$(v_{\rm rot}/\sigma_{\rm los})>1$ as ``disky''. }. For example, 
van Zee et al. (2004) found that 7/16 dEs in their sample showed dynamically
significant rotation. These authors used these data and other observables
to argue for multiple channels of dEs formation; ram-pressure stripping
of dwarf irregulars (dIs) in the galaxy cluster potential, tidal perturbation 
of dIs, and also for the infall of essentially pre-formed dEs.

However, a key problem in the kinematic analyses of dEs has been their
limited spatial extent. Deep integrated light observations generally
do not reach beyond one galaxy effective radius (R$_e$). For a disk galaxy
with a luminosity typical of luminous dEs, this roughly corresponds
to one disk scalelength. By contrast, HI rotation curves for spirals/dIs
may typically extend to beyond 5 disk scalelengths. If dEs were to lose mass and 
their stellar populations fade during their transformation in the cluster 
environment (e.g., Mastropietro et al. 2005) then the problem becomes
more acute; the more the progenitor galaxy has faded, the smaller fraction
of the progenitor disk scalelength can be measured in the remnant galaxy.

In this context, a powerful alternative to studying the integrated
stars of the galaxy itself is to study its globular clusters (GCs). The GC
systems of both giant and dwarf galaxies are more spatially extended
than the host galaxy light, reaching beyond 10 galaxy R$_e$ in some
cases, and with the appropriate instrumentation are 
amenable to spectroscopy (see Brodie \& Strader 2006 for a review).
In a pilot study, Beasley et al. (2006) obtained Keck/LRIS spectroscopy
for 12 GCs in the luminous Virgo dE VCC1087 extending out to $\sim5$
galaxy R$_e$. The kinematics of the 
GCs in this galaxy show $(v_{\rm rot}/\sigma_{\rm los})>1$, i.e. 
dynamically important rotation. The existing integrated
light data for this galaxy shows no rotation out to $\sim0.2$ R$_e$
(Geha et al. 2003). 

Clearly, it is hard to draw firm conclusions from one object.
Here we present GC kinematics for two more luminous Virgo dEs, VCC1261
and VCC1528, which show no morphological evidence of disk features (Lisker et al. 2006).
In Section 2 we discuss the observations and data reduction. In Section 3 we present
an analysis of the GC data, and also that of our integrated light measurements.
In Section 4 we discuss our findings.
In the following, we have assumed distance moduli to VCC1261 and VCC1528
of (m-M)=31.29 and 31.06 respectively, taken from the surface brightness 
distances of Mei et al. (2007). We also adopt $g$-band effective radii (R$_e$) of 
20.95 and 9.90 arcsec for the two dEs respectively (Ferrarese et al. 2006).




\section {Observations} \label{sec: 2}

Spectroscopic masks for VCC1261 and VCC1528 were 
created based on $g$ (F475W) and $z$ (F850LP)
images taken with the Advanced Camera for Surveys (ACS) as part of the 
ACS Virgo Cluster Survey (C{\^ o}t{\' e} et al.~2004).
A detailed discussion of our photometric reductions, photometry and 
candidate GC selection is given in Beasley et al. (2006) and 
Strader et al. (2006).

Optical spectra were obtained for 13 GC candidates in
VCC1261 and 12 candidates in VCC1528 using the 
Low Resolution Imaging Spectrograph (LRIS) (Oke et al.~1995) 
on the Keck I telescope during the nights 22-23 April 2007.
Both dEs were observed through a 1.0 arcsec slitmask
with integrations of $11\times1800$s for VCC1261 
and $7\times1800$s for VCC1528.
The locations of the GCs in the two galaxies are shown
in Fig. 1.
We obtained simultaneous blue and red spectra through the
use of a dichroic which split the beam at 5600\AA.
On the blue side, a 600 lines/mm grism blazed at 4000\AA\ was used 
to yield an effective wavelength range of 3300--5600\AA\ 
and a resolution of $\sim$4\AA\ (FWHM).
On the red side a 600 lines/mm grating blazed at 5000\AA\
yielded an effective wavelength range of 5700--8200\AA\ and
a resolution of $\sim$6\AA\ (FWHM).
Longslit spectra of both dEs ($3\times1200$s VCC1261; 
$6\times1200$s VCC1528) were also obtained along the major axes of these 
galaxies (P.A. 133 degrees and 97 degrees E through N 
for VCC1261 and VCC1528 respectively; Ferrarese et al. 2006). 
For the longslit observations, the blue-side setup was identical 
to that of the slitmasks, however on the red side, a 831 lines/mm 
grating was used (with a dichroic splitting the beam at 6800\AA)
blazed at 8200\AA\ to cover the Ca II triplet region. These Ca II
data will be discussed elsewhere.
Seeing ranged between 0.8--1.0 arcsec during observations 
of the VCC1261 mask on the first night, 
degrading somewhat to 1.1--1.3 arcsec for the VCC1528 mask on the second night.

The reduction of these data was performed with 
IRAF\footnote{IRAF is distributed by the National Optical Astronomy 
Observatory, which is operated by the Association of Universities for 
Research in Astronomy, Inc., under cooperative agreement with the 
National Science Foundation.} using standard techniques.
All science images were bias-subtracted and then flat-fielded with 
twilight sky flats which were co-added and normalized.
Spectra were optimally extracted and wavelength 
calibrated with solutions obtained from the arc exposures. 
Wavelength residuals of 0.1~\AA\ were typical. The zeropoints
of the wavelength calibration were checked against skylines in the 
background spectra, and were in some cases adjusted by up to 
an Angstrom. 
The spectra were then combined with cosmic-ray rejection, and flux
calibrated using the response function derived from our flux standard.
The velocities of the GC candidates were measured by cross-correlation
against stellar templates using FXCOR in IRAF. The zeropoints
of the velocity scale were checked against MILES model spectra 
(Vazdekis et al. 2008). 
These velocities, along with other relevant observables, are listed in Table~1.

\section {Analysis} \label{sec: 3}

We find robust mean velocities (i.e., the biweight location; Beers et al. 1990)
for the VCC1261 and VCC1528 GC systems of $1863\pm34$ \kms 
and $1689\pm28$ \kms respectively.
These are in agreement with the NED values for the systemic velocities 
of the two galaxies ($1871\pm16$ \kms and $1647\pm25$ \kms). 
Using the biweight scale as a robust measure of the line-of-sight velocity 
dispersions of the GC systems, we obtain $67\pm25$ \kms and $50\pm21$ \kms 
for VCC1261 and VCC1528 respectively.  These quantities are listed in Table~2.
For completeness, we also tabulate the corresponding quantities for VCC1087, either
listed in, or derived from Beasley et al. (2006).

We looked for rotation in the GC systems using two approaches. 
We first performed linear unweighted fits to the
velocities of the GCs in the two galaxies as a function of 
projected distance along the galaxy major axes. 
This assumes that any rotation that might be present in the system 
is solid body rotation, and that it has its axis is aligned with the 
galaxy minor axis. 
Because of our relatively small sample size and 
wide range in velocity uncertainties, we avoided using weights 
since the fits might be driven by a few bright GCs
(although in practice we found little difference
between weighted and unweighted fits). The results of this exercise are 
shown in the top panels of Fig. 2, and are listed in Table 2. 
Both GC systems exhibit velocity gradients as a function of project radius
along the galaxy major axis, suggestive of net rotation. A ``maximum velocity''
($v_{\rm max}$) was calculated in both cases by multiplying the velocity 
gradient with the projected radius of the outermost GC. 
We calculated two values for VCC1261, with and without GC 250 which 
appears to be an outlier (or is no longer on the rising part of the rotation 
curve if we are actually looking at a disk system)\footnote{Also, two values
are given for VCC1087 including and excluding GC 2, which due to its
large galactocentric radius, has a strong influence on the fit to these velocities
(see Beasley et al. 2006).}.

We estimated the significance of this rotation through Monte Carlo simulation.
We randomly selected $N$ velocities from a Gaussian probability distribution of 
width equal to the galaxies' velocity dispersions (where $N$ corresponds to our 
GC sample size). We then performed linear fits to these velocities
as a function of radius, which were drawn randomly from the observed radial 
extent of the GC systems, and asked how many times these simulated gradients 
were found to be equal to or larger than the observed ones. For VCC1261 we 
obtained significances of 85\% (97\%) including (excluding) GC250. 
In the case of VCC1528, the observed
rotation is significant at the 96\% level.

The above considerations assume that the line of nodes of any rotation
in the GC systems is aligned with the galaxy major axes.
To test this assumption we performed non-linear fits to the GC 
velocities as a function of their position angle with respect to the 
galaxy. Specifically, we fit sinusoids (see in discussion C{\^ o}t{\' e} et al.~2001)
to the position angle data in the form:

\begin{equation} \label{eq:one}
v(\theta)=v_{\rm sys} + v_{\rm rot} ~ {\rm sin}(\theta-\theta_{\rm sys})
\end{equation}

\noindent leaving $v_{\rm sys}$, $v_{\rm rot}$, and $\theta_{\rm sys}$ 
(the systemic velocity, rotation amplitude and phase -- 
the position angle of the line of nodes -- respectively) as free parameters.
The results of these fits are listed in Table 2, and are 
shown as dotted sinusoids in the bottom panels of Fig. 2.
As shown in the table, the amplitude of the rotation in the GC systems
is consistent with the values derived from the linear fits.
The solid sinusoids in Fig. 2 show the best fits for the amplitude
and systemic velocities when the phase is fixed to coincide with 
the position angles of the galaxy major axes. 
In the case of VCC1261, the two curves coincide
suggesting that the axis of rotation is indeed close to the minor 
axis of the galaxy. For VCC1528 the situation is less clear;
the rotation could be misaligned with the galaxy isophotes, but
the large uncertainty in our phase estimate prevents any firm 
conclusions from being drawn.

A common measure of the significance of rotation in a system is the 
ratio of rotational velocity to velocity dispersion ($v_{\rm rot}/\sigma_{\rm los}$).
Here we take $v_{\rm rot}$ to be $v_{\rm max}$,  derived from our earlier 
linear fits to the GC velocities as a function of projected major axis radius.
We then subtract a rotation curve of amplitude $v_{\rm max}$ from 
the observed velocity dispersions of the GC systems in order to measure 
$\sigma_{\rm los}$. These quantities are listed in Table~2. The GC 
systems of both dEs show $v_{\rm rot}/\sigma_{\rm los} >1$, which 
suggests dynamically significant rotation.

\subsection{Galaxy kinematics}

In Fig. 3 we show the major-axis kinematics of the 
two galaxies from our longslit data\footnote{At the time
of writing, we do not have deep longslit data for VCC1087.}. We reach
approximately 1 R$_e$ in VCC1261 and 2 R$_e$ in VCC1528.
Both galaxies show fairly complex kinematical structure. Both show
significant velocity gradients at small radii, which seem to become flatter
towards the outer parts of the galaxies. 
Similarly, the galaxy velocity dispersions are lower at radii corresponding to the 
velocity gradients, increasing at larger radii. 
To give a sense of scale, 1 R$_e$ in VCC1261 and VCC1528 corresponds to 
approximately 1 and 2 kpc respectively at our adopted distances. Thus, these 
inner structures have spatial scales of $\sim0.5$ kpc.
Such kinematical structures have been noted before in dEs (De Rijcke et al. 2004; 
Geha et al. 2006) and have been coined kinematically distinct cores (KDCs)
in deference to the similar structures seen in giant ellipticals.

The dotted lines in Fig. 3 show the velocity 
gradients derived for the GCs. The GC and galaxy inner gradients are
similar in both magnitude and sign; this is most apparent in the 
zoomed-in center panels of Fig. 3.
Quantifying this, by eye we isolated the galaxy inner regions and performed linear fits. 
For VCC1261 we obtain a gradient of --52.9$\pm7.1$ \kms arcmin$^{-1}$ ($-0.2<$ R$_e$ $<0.2$).
For VCC1528 we obtain 62.9$\pm6.3$ \kms arcmin$^{-1}$ ($-0.4<$ R$_e$ $<0.4$).
Linear fits to the velocity data at all radii yields 2.6$\pm4.5$ \kms arcmin$^{-1}$
and --6.3$\pm8.3$ \kms arcmin$^{-1}$ for VCC1261 and VCC1528 respectively.

However, in general, for systems where the velocity dispersion is of order the rotation velocity, 
the observed rotation velocity will not reflect the true circular velocity of the system 
since the velocity dispersion provides additional dynamical support against gravity.
We make such ``asymmetric drift'' corrections to the observed galaxy rotation
curves using the following (e.g., Kormendy 1984):

\begin{equation}
V^2_{\rm c}(r)=V^2_{\rm obs}(r)+\sigma^2_{\rm R}(r)(2r\alpha-1)
\end{equation}

\noindent where $V_{\rm c}(r)$ is the corrected rotation curve at radius
$r$, $V_{\rm obs}$ the observed rotation, and $\sigma_{\rm R}$ the radial 
component of the galaxy velocity dispersion. This latter quantity is not 
measured by us, and we make the assumption that
$\sigma_{\rm R}=\sqrt{2}\sigma_{\rm obs}$, with $\sigma_{\rm obs}$ being 
the measured galaxy velocity dispersion (e.g., see Erwin et al. 2003). $\alpha$ is the 
inverse of the disk scalelength, which we set to be the galaxy effective radius.

The observed and corrected rotation curves for the dEs are compared in
Fig. 4. As can be seen the ``asymmetric drift'' corrections are substantial\footnote{The asymmetric drift corrections are computed assuming 
a pure thin disk. Since the velocity dispersions in the galaxy are significant, it may be more appropriate to assume a thick disk, or a thin disk plus ``hot''  component. In these two cases the corrections would be substantially reduced. However, since we cannot distinguish between these posibilities with the present data, we leave  these corrections ``as is''.}.

Weak or no rotation in the observed curves becomes significant rotation in the 
corrected curves when the dynamical role of velocity dispersion is
accounted for. Therefore, both dEs show strong {\it global} rotation in 
the opposite sense to the rotation seen in the GC systems.


\subsection{Mass to light ratios}

We have estimated the dynamical masses of the dEs by considering that the
total mass of the system ($M_{\rm tot}$) as traced by the GCs is given by:

\begin{equation}
M_{\rm tot}=M_{\rm pres}+M_{\rm rot}
\end{equation}

\noindent where $M_{\rm pres}$ and $M_{\rm rot}$ are the masses inferred from 
the velocity dispersion (after subtracting the corresponding rotation curve) and rotation of the GCs respectively. 
$M_{\rm pres}$ can be estimated using the tracer mass estimator (TME; Evans et al. 2003):

\begin{equation}
M_{\rm pres}=\frac{C}{GN}\sum_i v^2_i R_i
\label{TME}
\end{equation}

\noindent where $N$ is the number of test particles (in this case GCs), 
$v_i$ is the line-of-sight velocity of the $i$th GC and 
$R_i$ is the projected distance of the $i$th GC
from the center of the galaxy. 
Assuming an isothermal-like potential, $C$ is given by (Evans et al.~2003):

\begin{equation}
C=\frac{4\gamma}{\pi}\frac{4-\gamma}{3-\gamma}\frac{1-(r_{\rm in}/r_{\rm out})^{3-\gamma}}
{1-(r_{\rm in}/r_{\rm out})^{4-\gamma}}
\label{TME2}
\end{equation}

\noindent here, $r_{\rm in}$ and $r_{\rm out}$ are the inner and outer radii at 
which the volume density of the population goes as $r^{-\gamma}$.
We measured the surface densities of the GC systems from the ACS photometry, 
fitting the radially binned data with the functional form $\rho(r) \propto r^{-\alpha}$. 
These were then deprojected to yield $\gamma$ (i.e., $\alpha+1$).

The rotational mass, $M_{\rm rot}$, was calculated assuming solid-body rotation using:

\begin{equation}
M_{\rm rot} =\frac{r_{\rm out}v^2_{\rm max}}{G}
\end{equation}

We list our measurements of $\gamma$, the dynamical masses and the B-band
mass-to-light ratios ($\Upsilon_B$) in Table 2.
The B-band absolute magnitudes for VCC1261 and VCC1528 
were derived from the apparent magnitudes in Binggeli et al. (1985), the
distance moduli from Mei et al. (2006) and foreground extinction values from 
Schlegel et al. (1998).

In Fig 5. we plot the $\Upsilon_B$ derived for VCC1261, VCC1528 and VCC1087
compared to determinations from integrated light kinematics for Elliptical and S0 galaxies 
($M_{\rm Jean}$ in Cappellari et al. 2006) and Virgo dEs (Geha et al. 2002). 
Unlike the case for our dE sample, the above studies obtain dynamical masses 
for the galaxies inside an R$_e$. 
We have not included predictions from stellar population models, since these are
highly dependent upon the choice of initial mass function, and the adopted lower stellar
mass cut-off. Using GCs to increase the radius at which
mass can be traced, we see that the three dEs in our sample lie above the 
measurements from Cappellari et al. (2006) and Geha et al. (2002) for a given 
B-band magnitude. However, as can be seen from our estimates of $M_{\rm press}$ and $M_{\rm rot}$
listed in Table 2, $M_{\rm rot}$ dominates the mass budget at large radii.
In the cases of VCC1261 and VCC1528, large $M_{\rm rot}$ values are driven by 
single GCs. Excluding these objects brings $M_{\rm rot}\simeq M_{\rm press}$
and reduces $\Upsilon_B$ somewhat. 

In our estimates of the rotation of the GCs, we do not take into account the 
inclination of the GC systems due to the unknown the geometry of the 
systems. However, in the extreme case that we are viewing what are 
essentially thin disks at some unknown inclination $i$, and that the GC systems are 
inclined similarly to the host dE isophotes, we may estimate 
the inclinations of the systems (i.e., sin $i$=1--$\epsilon$). In this case, 
the GC systems of VCC1261 and VCC1528 may be inclined (from edge-on, $i=90\degr$)
by $45\degr$ and $56\degr$ respectively. This being the case, our values
for $v_{\rm max}$ for the GC systems of VCC1261 and VCC1528 would be underestimated 
by 70\% and 83\% respectively, and the mass to light ratios we have determined
would be underestimated by a factor of $\sim$2.5 for both systems.

\section {Discussion and Summary} \label{sec: 4}

The kinematics of the three GC systems of luminous Virgo dwarf elliptical 
galaxies studied in detail to date, VCC1261, VCC1528 and VCC1087 (Beasley et al. 2006), 
all show evidence for significant rotation, i.e. $(v_{\rm rot}/\sigma_{\rm los})>1$.
For these GC systems to exhibit such strong rotation, they must have either 
acquired their angular momentum within the cluster environment (i.e. they were ``spun-up''), 
and/or their progenitors were already rapidly rotating systems. 

De Rijcke et al. (2004) discuss the former possibility 
within the context of explaining the presence of kinematically decoupled
cores in the centers of some dEs. These authors used the impulse approximation
to estimate the amount of orbital angular momentum which may be transferred to a
dE halo from fly-by encounters with massive galaxies. 
These authors conclude that such encounters can spin-up slowly rotating systems by 
tens of \kms for small impact parameters ($\leq$ 20 kpc). However, $N$-body simulations by 
Gonz\'alez-Garc\'ia et al. (2005) suggest that this process is much less
efficient than the impulse approximation suggests. They found that essentially
head-on collisions are required to achieve significant velocity reversals (in 
already rotating systems). The process becomes even less efficient if the dE
possesses a dark matter halo, since the halo absorbs much of the orbital angular 
momentum of the encounter. Numerical simulations of fly-by encounters
between a massive galaxy and dEs+GCs in the cluster environment are required to 
investigate this possibility further.

Another interpretation is that the dEs were
formerly late-type galaxies, and their GCs formed part of what was a disk 
component. Indeed, many dEs show embedded disks (Barrazza et al. 2002; 
De Rijcke et al. 2003; Lisker et al. 2006; Chilingarian et al. 2007), 
or rotational flattening of the spheroid stars 
(Pedraz et al. 2002; Geha et al. 2003; van Zee et al. 2004). 
$N$-body simulations suggest that galaxy cluster tidal forces can strip stars, gas and dark matter 
from a late-type disk galaxy that falls into the cluster potential (Moore et al. 1996). 
Mastropietro et al. (2005) showed that subsequent internal dynamical processes
can transform the stripped disk into a hot spheroidal system on Gy timescales, 
leading to a remnant that looks like a dE galaxy. 
Importantly, the majority of simulated dEs
formed in this manner exhibit some memory of their previous condition,  
morphologically (residual disk structure) and/or kinematically (rotation).

The idea that old GCs in late-type 
galaxies may belong to disk rather than halo populations finds support in the work of 
Olsen et al. (2004). These authors found that the GC systems of Sculptor 
spirals show rotation broadly consistent with their HI gas. 
A similar situation may also be true of the GCs in the Large Magellanic Cloud 
(Schommer et al. 1992) and in M31 (Perrett et al. 2002; Morrison et al. 2005)
(although in the latter case sample contamination confuses the issue; 
Beasley et al. 2004; Cohen et al. 2005). Interestingly, the GC systems of several 
S0 galaxies also show disk-like kinematics (Kuntschner et al. 2002; Chomiuk et al. 2008).

One expectation of the above interpretation is that 
the progenitor disk should have obeyed the Tully-Fisher (TF) relation 
(Tully \& Fisher 1977).
In Fig. 6 we show the $r$-band TF relation for isolated galaxies from
Blanton et al. (2007). On this we have placed the three VCC galaxies, using 
SDSS $r$-band luminosities and the SBF distances of Mei et al. (2007). 
The rotation velocities ($v_{\rm max}$) 
are our measurements of the GC systems and from Beasley et al. (2006). 
The positions of all three dwarfs are consistent with our fit to the 
three brightest TF bins ($M_r$--5 log$~h<-18$; $h=0.7$). 

If we are looking at a remnant population, one might expect
that the progenitor disks were more luminous than the remnants due to
mass-loss during ``harassment'', and due to the subsequent ageing of the 
stellar populations.
The arrows in Fig. 6 show the maximum brightening allowed for the dEs to
remain consistent with the fit to the relation. 
These correspond to 1--2 mag depending upon
the galaxy. An extra 0.5 mag is permitted if the scatter in the TF relation 
is also taken into account. In terms of stellar mass-loss, this corresponds 
to $\sim$70--95\% stellar mass loss for a given mass to light ratio, which is 
consistent with the levels of mass-loss seen in the simulations of 
Mastropietro et al. (2005) for dwarfs near the cluster center. 
We have not included age-fading which will reduce the
amount of allowable mass-loss somewhat. The point here is that any 
progenitor disks could have been up to 2 mag brighter than the remnant dEs 
and still lie on the TF relation. In contrast, in the previously discussed
scenario whereby the dEs are spun-up by fly-by encounters, one would not expect 
these galaxies to lie on the TF relation. Clearly, decreasing the uncertainties 
in our rotation estimates through increasing the sample sizes and/or reducing 
the individual velocity uncertainties would help constrain these ideas.

In terms of the galaxy kinematics, VCC1261 and VCC1528 show corotation similar to the 
GCs within $\sim0.4$ R$_e$, with weak or no (counter)rotation beyond this radius.
The situation for VCC1087 is unclear since we do not possess deep longslit 
data for this galaxy. The similarity between the inner galaxy kinematics and the 
GCs may be coincidental; the GC rotation is measured out to $\sim5$ R$_e$ whereas the 
``inner disks'' do not extend beyond $\sim0.4$ R$_e$. However, the fact that this 
behaviour is seen in both dEs seems to suggest a real connection between the two.

Irrespective of whether the galaxy inner regions and the GCs are kinematically
connected, more puzzling is why the galaxy kinematics beyond $\sim0.5$ R$_e$ show
{\it no} rotation. It is possible that the GCs and the galaxy stars
represent distinct kinematical components. Since any external dynamical
process (such as fly-by encounters) will presumably affect the stars and GCs similarly
at a given radius, a kinematical decoupling argues for some level of dissipation. 
For example, we may speculate that a dE with gas and a pre-formed GC disk system 
could have fallen into the cluster. This gas may then have been shocked and formed
stars, resulting in a system with an old disk and a younger pressure-supported system.

Although the possibility of such decoupling between stars and GCs 
cannot presently be ruled out, a more prosaic explanation may be 
that the dE stars are rotating, but we cannot see it due to the limited radial extent 
of these galaxy data.
Our longslit data for VCC1261 and VCC1528 reaches $\sim1$ and $\sim2$
R$_e$ respectively (1.6 kpc and 0.8 kpc respectively, 
typical of deep, longslit observations). 
If these galaxies were dwarf irregulars with the luminosities of the 
present-day dEs, these data would correspondingly reach to $\sim$1--2 disk 
scalelengths. 
By way of comparison, typical HI rotation curves extend to $\sim5$ disk 
scalelengths (see van Zee et al. 2004). 

Since disk galaxies obey a well-defined luminosity-size relation, 
this lack of radial coverage in the longslit data of dEs becomes more acute 
if we consider that the hypothetical progenitor disks may have been more luminous 
than their remnants.
If, for example, we consider a disk galaxy 1 magnitude more luminous than 
VCC1261, which may have had a disk scalelength of $\sim5$ kpc, we find 
that our longslit data might be probing a mere $\sim0.3$ disk scalelengths of 
this galaxy. A similar argument for VCC1528 suggest we might probe $\sim0.5$
disk scalelengths. In contrast, the outermost GCs in our two samples lie at 
$\sim4$ and $\sim7$ R$_e$ (6.4 kpc and 5.6 kpc) for VCC1261 and
VCC1528 respectively, corresponding to 1.3 and 3.5 disk scalelengths. 
Therefore, it seems quite possible that the kinematical data in these galaxies 
simply does not go out sufficiently far in radius to be 
sensitive any rotation present. 

Clearly detailed numerical simulations are required to address some of the above
ideas. However, based on these data we argue that the GC systems of luminous dEs 
may represent vestigal disk systems of late type galaxies. The identification
of disk GC systems in both the Virgo cluster and in the Sculptor group (Olsen et al. 2004) 
suggests that gas disks may have been significant sites of GC formation 
at early times.

\section{Acknowledgments}

MB and AJC thank Inma Mart\'inez Valpuesta for very insightful discussions about disks.
The authors also thank the anonymous referee for useful suggestions on improving the 
manuscript.
Funding for the SDSS and SDSS-II has been provided by the Alfred P. Sloan Foundation, the 
Participating Institutions, the National Science Foundation, the U.S. Department of Energy, 
the National Aeronautics and Space Administration, the Japanese Monbukagakusho, the Max Planck 
Society, and the Higher Education Funding Council for England. 
The SDSS Web Site is http://www.sdss.org/. The SDSS is managed by the Astrophysical Research 
Consortium for the Participating Institutions. The Participating Institutions are the American 
Museum of Natural History, Astrophysical Institute Potsdam, University of Basel, University of 
Cambridge, Case Western Reserve University, University of Chicago, Drexel University, Fermilab,
the Institute for Advanced Study, the Japan Participation Group, Johns Hopkins University, 
the Joint Institute for Nuclear Astrophysics, the Kavli Institute for Particle Astrophysics 
and Cosmology, the Korean Scientist Group, the Chinese Academy of Sciences (LAMOST), 
Los Alamos National Laboratory, the Max-Planck-Institute for Astronomy (MPIA), the
Max-Planck-Institute for Astrophysics (MPA), New Mexico State University, Ohio 
State University, University of Pittsburgh, University of Portsmouth, Princeton University, 
the United States Naval Observatory, and the University of Washington. 

This research has made use of the NASA/IPAC Extragalactic Database (NED) which is 
operated by the Jet Propulsion Laboratory, California Institute of Technology, under 
contract with the National Aeronautics and Space Administration.

Funding support comes from NSF grant AST-0507729.
J.S. was supported by NASA through a 
Hubble Fellowship grant, awarded by the Space Telescope Science Institute,
which is operated by the Association of Universities for Research in
Astronomy, Inc., under NASA contract NAS 5-26555.
The authors wish to recognize and acknowledge the very significant cultural role and 
reverence that the summit of Mauna Kea has always had within the indigenous Hawaiian 
community. We are most fortunate to have the opportunity to conduct observations from 
this mountain.

{}

\clearpage
\begin{figure}
\vspace{-4.0 cm}
\hspace{-2.5 cm}
\epsscale{1.0}
\plotone{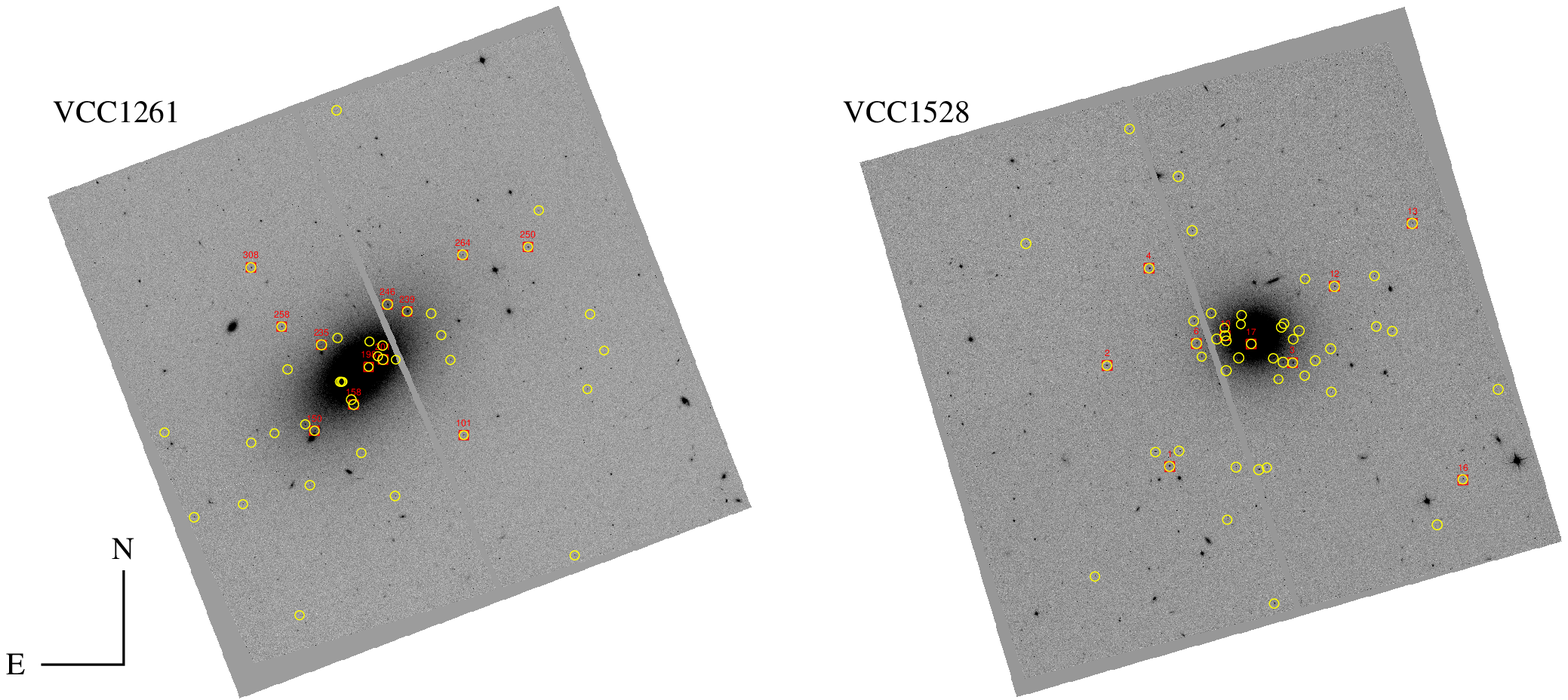}
\caption{Spatial distribution of GCs in VCC1261 and VCC1528 marked on the ACS $z$ images 
(3.5 arcmin on a side). Circles represent GCs detected in the $g$ and $z$ ACS images. 
The numbered circle-boxes are those GCs for which we obtained spectra.
}
\end{figure}

\clearpage
\begin{figure}
\vspace{-2.0 cm}
\hspace{-0.5 cm}
\epsscale{0.8}
\plotone{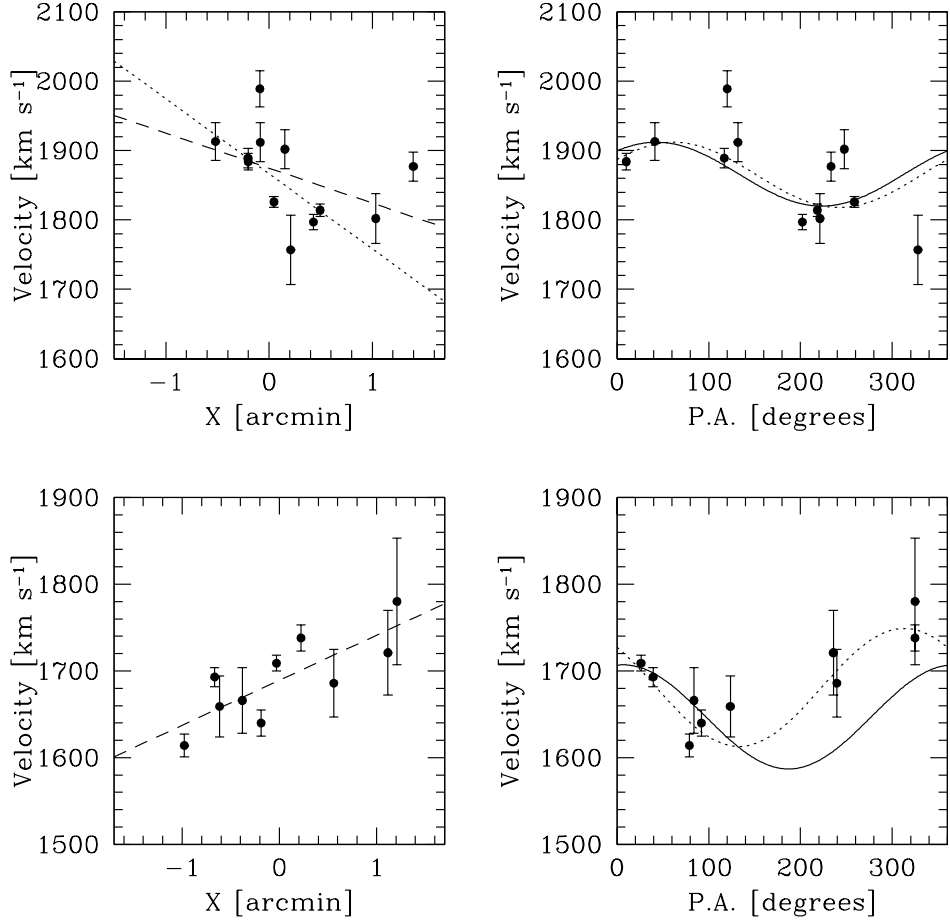}
\caption{Kinematics of VCC1261 GCs ({\it top panels})
and VCC1528 GCs ({\it bottom panels}). The left-hand panels
show weighted least-squares linear fits (dotted lines) to the 
velocities as a function of projected distance along the major axes of 
the galaxies. The dotted line shows the fit for VCC1261 excluding 
the outermost GC (GC250).
The right-hand panels show the GC velocities versus azimuthal position
(measured E through N). The dotted lines are best-fit sinusoids leaving the 
systemic velocities, position angles of the line of nodes and amplitudes as free 
parameters. The solid lines are sinusoidal fits fixing the position angle
to coincide with that of the galaxy major axes.
}
\end{figure}

\clearpage
\begin{figure}
\vspace{-2.0 cm}
\hspace{-0.5 cm}
\epsscale{1.0}
\plotone{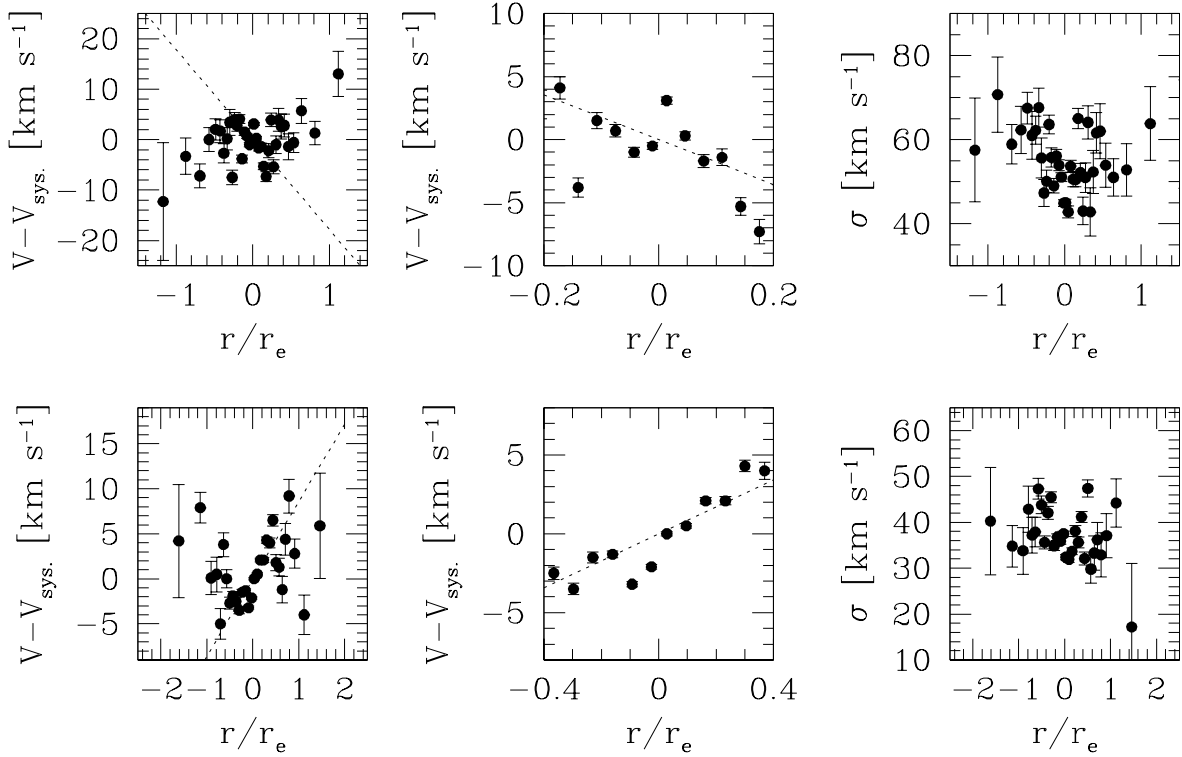}
\caption{Kinematics of VCC1261 ({\it top panels})
and VCC1528 ({\it bottom panels}). 
{\it Left panels:} the mean velocities of the integrated
galaxy light (correct for the systemic velocity) as a function of projected 
radius along the major axis. The dotted lines are the velocity gradients
derived for the GCs, and are {\it not} fits to the galaxy data. 
{\it Center panels:} zoomed-in versions of the left-hand panels comparing 
the central galaxy rotation to that of the GCs. Note the change of scale. 
{\it Right panels:} velocity dispersion of 
integrated galaxy light as a function of projected radius.
All individual points represent radial bins such that the S/N per
bin is $\geq$15 \AA$^{-1}$.
}
\end{figure}

\clearpage
\begin{figure}
\vspace{-2.0 cm}
\hspace{-0.5 cm}
\epsscale{1.0}
\plotone{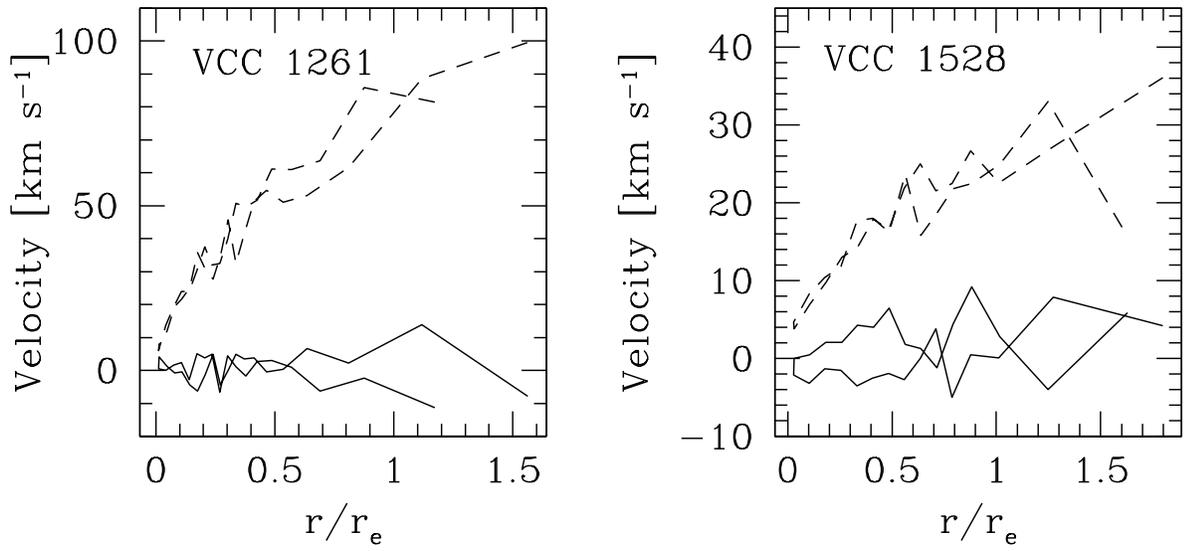}
\caption{Observed rotation ({\it solid lines}) and rotation curves
corrected for asymmetric drift ({\it dashed lines}) for VCC1261 and VCC1528.}
\end{figure}

\clearpage
\begin{figure}
\vspace{-2.0 cm}
\hspace{-0.5 cm}
\epsscale{1.0}
\plotone{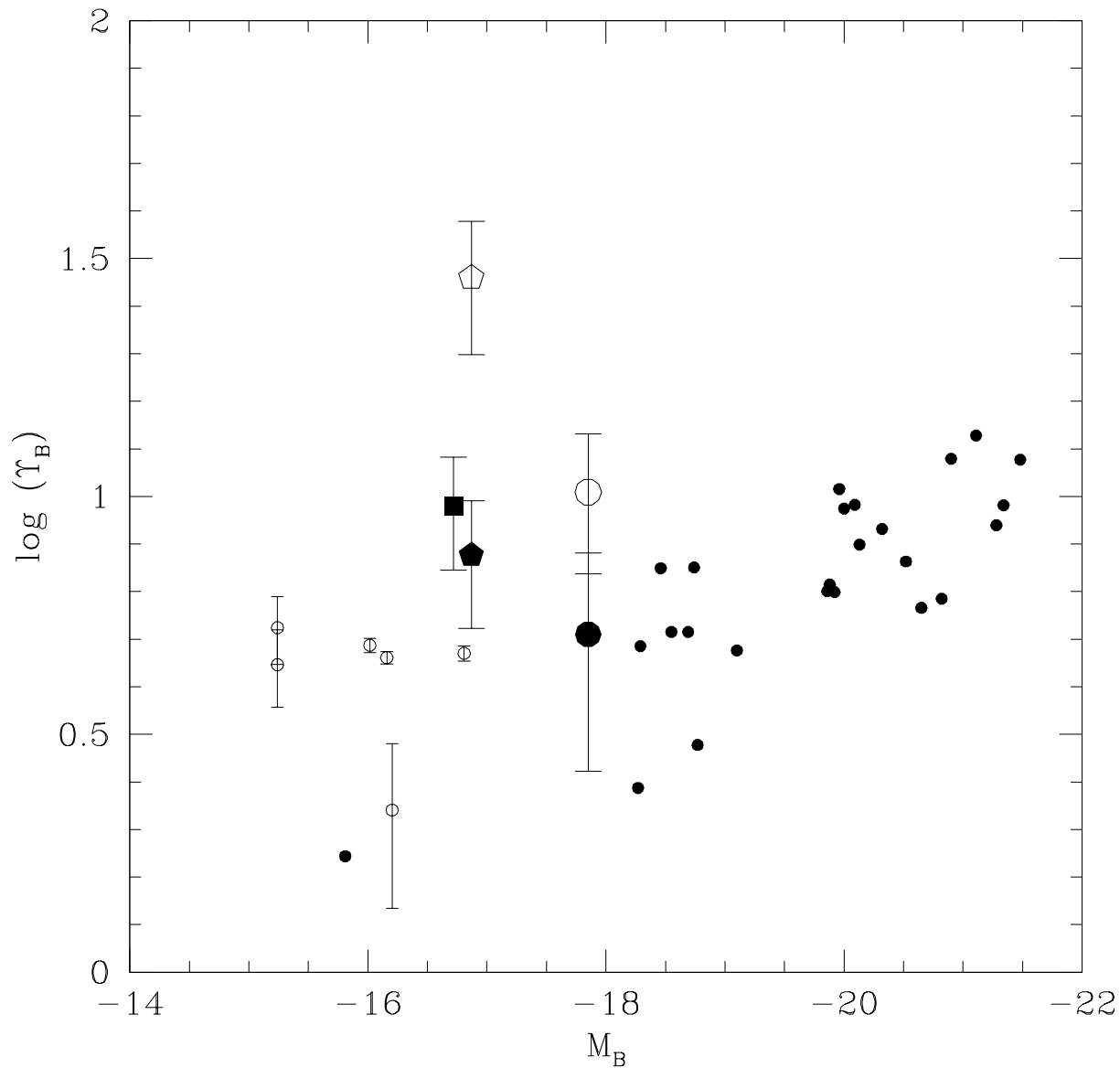}
\caption{B-band mass to light ratios of the dEs calculated using the 
TME ({\it square:} VCC1528;  {\it pentagons:} VCC1087; {\it circles:} VCC1261)
compared to the estimates for E and S0 galaxies of SAURON galaxies 
({\it small solid circles}; Cappellari et al. 2006) and Virgo dEs (Geha 
et al. 2003). The open and solid symbols for VCC1087 and VCC1261 show masses
calculated including and excluding the outermost GC respectively, where the mass
contribution is dominated by solid-body rotation.}
\end{figure}

\clearpage
\begin{figure}
\vspace{-2.0 cm}
\hspace{-0.5 cm}
\epsscale{0.8}
\plotone{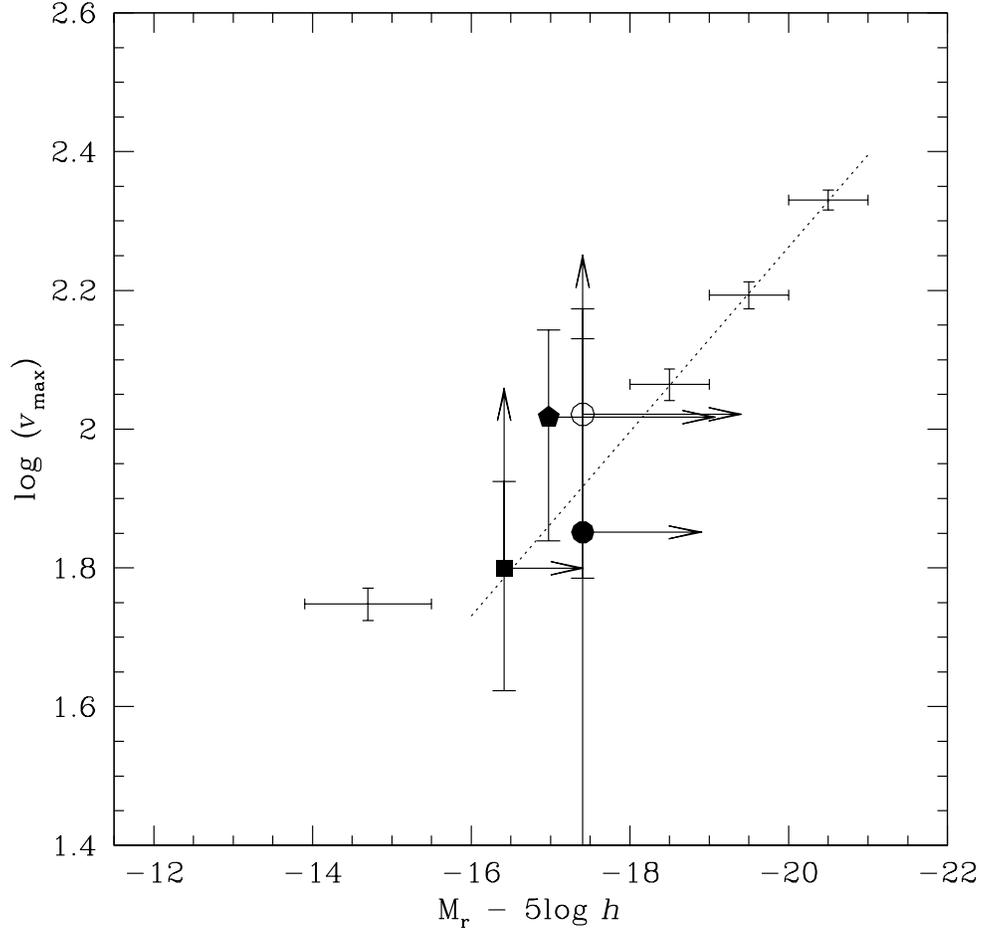}
\caption{Positions of dwarf ellipticals, based on the rotation
of their GC systems, in comparison to the $r$-band Tully-Fisher relation for 
isolated galaxies. Crosses show the median values for $v_{\rm max}$ for isolated galaxies
taken from Blanton et al. (2007), based on data from Geha et al. (2006), 
Pizagno et al. (2006) and Springob et al. (2005). The dotted line is our
least-squares fit to the bins with $M_r$--5 log$~h<-18$.
VCC1261, VCC1528 and VCC1087 (from Beasley et al. 2006) are
shown as circles, the solid square and the solid pentagon respectively.
The open and solid circles represent our two measurements of $v_{\rm max}$
for VCC1261. Vertical arrows demonstrate the maximum effect of correcting 
for unkown inclination of the GC systems, horizontal arrows show the maximum 
allowable ``unfaded'' magnitudes
of the dEs which are consistent with the linear fit (see text).
}
\end{figure}

\begin{deluxetable}{lcccclc}
\tablecolumns{6}
\tablewidth{0pc}
\tablecaption{Data for VCC1261 and VCC1528}
\tablehead{
\colhead{Target} & \colhead{RA(J2000)} & \colhead{DEC(J2000)} & \colhead{$g$} 
& \colhead{$g-z$} & \colhead{RV} & \colhead{Class.} \\
\colhead{} & \colhead{} & \colhead{} & \colhead{(AB mag)} & \colhead{} &
\colhead{(\kms)} & \colhead{}\\
}
\startdata 
VCC1261 & & & & & & \\
\hline
GC250 & 12:30:05.98 & 10:47:34.76 & 22.35$\pm$0.01 & 1.03$\pm$0.01 & 1877$\pm$21 & GC\\
GC101 & 12:30:07.59 & 10:46:20.50 & 23.55$\pm$0.02 & 0.77$\pm$0.03 & 1757$\pm$50 & GC\\
GC264 & 12:30:07.69 & 10:47:31.26 & 23.74$\pm$0.02 & 1.14$\pm$0.03 & 1802$\pm$36 & GC\\
GC239 & 12:30:09.13 & 10:47:08.76 & 21.65$\pm$0.01 & 1.17$\pm$0.01 & 1814$\pm$9  & GC\\
GC246 & 12:30:09.65 & 10:47:11.39 & 21.88$\pm$0.01 & 1.20$\pm$0.01 & 1797$\pm$11 & GC\\
GC201 & 12:30:09.74 & 10:46:49.73 & 22.54$\pm$0.02 & 1.07$\pm$0.03 & 1902$\pm$28 & GC\\
GC198 & 12:30:10.12 & 10:46:46.72 & 23.29$\pm$0.13 & 1.07$\pm$0.20 & 1826$\pm$8  & GC\\
GC158 & 12:30:10.50 & 10:46:31.86 & 21.84$\pm$0.01 & 0.96$\pm$0.01 & 1884$\pm$12 & GC\\
GC235 & 12:30:11.37 & 10:46:55.13 & 23.51$\pm$0.02 & 0.96$\pm$0.03 & 1989$\pm$26 & GC\\
GC150 & 12:30:11.52 & 10:46:21.37 & 23.37$\pm$0.03 & 0.97$\pm$0.04 & 1913$\pm$27 & GC\\
GC258 & 12:30:12.42 & 10:47:02.05 & 22.64$\pm$0.01 & 0.89$\pm$0.01 & 1889$\pm$14 & GC\\
GC308 & 12:30:13.25 & 10:47:25.12 & 22.52$\pm$0.01 & 1.03$\pm$0.02 & 1912$\pm$28 & GC\\
GC113 & 12:30:13.37 & 10:45:52.11 & 23.77$\pm$0.02 & 0.76$\pm$0.03 &  50$\pm$29 & star\\
\hline
VCC1528 & & & & & & \\
\hline
GC16 & 12:33:46.35 & 13:18:25.6 & 23.60$\pm$0.02 & 0.75$\pm$0.04 & 1780$\pm$73 & GC\\
GC13$^a$ & 12:33:47.53 & 13:20:02.5 & 22.90$\pm$0.20 & 1.60$\pm$0.40 & 1721$\pm$49 & GC\\
GC8  & 12:33:48.73 & 13:21:06.9 & 23.07$\pm$0.01 & 0.85$\pm$0.02 &  85$\pm$29  & star\\
GC12 & 12:33:49.54 & 13:19:39.2 & 23.02$\pm$0.01 & 0.77$\pm$0.02 & 1686$\pm$39 & GC\\
GC3  & 12:33:50.64 & 13:19:10.8 & 22.08$\pm$0.01 & 0.82$\pm$0.01 & 1738$\pm$15 & GC\\
GC17 & 12:33:51.69 & 13:19:18.1 & 22.91$\pm$0.06 & 1.08$\pm$0.10 & 1709$\pm$9  & GC\\
GC10 & 12:33:52.35 & 13:19:21.4 & 23.05$\pm$0.03 & 0.90$\pm$0.05 & 1640$\pm$15 & GC\\
GC6  & 12:33:53.09 & 13:19:18.7 & 22.71$\pm$0.01 & 0.92$\pm$0.02 & 1666$\pm$38 & GC\\
GC1  & 12:33:53.83 & 13:18:32.5 & 21.72$\pm$0.01 & 0.80$\pm$0.01 & 1693$\pm$11 & GC\\
GC4  & 12:33:54.26 & 13:19:47.4 & 22.39$\pm$0.01 & 0.94$\pm$0.01 & 1659$\pm$35 & GC\\
GC2  & 12:33:55.38 & 13:19:11.0 & 22.07$\pm$0.01 & 0.96$\pm$0.01 & 1614$\pm$13 & GC\\
GC11 & 12:33:57.81 & 13:18:21.4 & 23.00$\pm$0.01 & 1.10$\pm$0.02 & 239$\pm$24  & star\\
\enddata
\tablenotetext{a}{Object lies on chip defect in ACS. Photometry taken from SDSS.}
\end{deluxetable}

\begin{deluxetable}{llccc}
\tablecolumns{3}
\tablewidth{0pc}
\tablecaption{Kinematical properties of GC systems in Virgo dEs}
\tablehead{
\colhead{Quantity} & \colhead{} & \colhead{VCC1261} & \colhead{VCC1528} & \colhead{VCC1087}\\
}
\startdata 
M$_{\rm B}$	&  	& --17.85  		& --16.72 & --16.87\\
$v_{\rm grad}$ &  (\kms arcmin$^{-1}$) & --51$\pm$46 (--101$\pm$42) & 52$\pm$17 & 87$\pm$29 (74$\pm$31)\\
$v_{\rm max}$ &  (\kms)  &  71$\pm$64 (105$\pm$44) & 63$\pm$21 & 104$\pm$35 (49$\pm$21)\\
$v_{\rm rot}$ & 	(\kms)	& 47$\pm$31 	& 68$\pm$40\\
$\theta_{\rm sys}$ & (degrees) & 152$\pm$27 	& 43$\pm$136 & 129$\pm$50\\
$v_{\rm sys}$ & (\kms)  & 1865$\pm$19 	& 1681$\pm$21 & 681$\pm$19\\
$\sigma_{\rm meas.}$ & (\kms)	& 67$\pm$25 	& 50$\pm$21 & $49\pm$15 \\ 
$\sigma_{\rm los}$ &  (\kms)	& 55$\pm$13 & 23$\pm$9 & 35$\pm$16$^{a}$\\
$(v_{\rm rot}/\sigma_{\rm los})$ & & 1.3$\pm$0.7 (1.8$\pm$0.5)  & 2.7$\pm$0.5 & 3.0$\pm$0.6$^{b}$ (1.6$\pm$0.6)\\
$\gamma$ &			& 1.9$\pm$0.4 & 2.0$\pm$0.3 & 2.1$\pm$0.4\\
M$_{\rm pres}$ & ($10^{10}$\msun) 		& 0.52$\pm$0.17 (0.53$\pm$0.16)  & 0.20$\pm$0.09 & 0.35$\pm$0.15 (0.49$\pm$0.23)\\
M$_{\rm rot}$ & ($10^{10}$\msun)		& 0.56$\pm$0.50 (1.67$\pm$0.7)  & 0.52$\pm$0.17 & 2.21$\pm$0.64 (0.34$\pm$0.15)\\
$\Upsilon_B$ &	& 5.0$\pm$2.5 (10.2$\pm$3.3)  & 9.5$\pm$2.5 & 29.3$\pm$8.7 (6.8$\pm$2.2)\\ 
\enddata
\tablecomments{Rows denote (1): absolute B-band magnitude, (2) velocity gradient from
linear fits, (3) maximal rotation from linear fits, (4) rotation from nonlinear fits,  
(5) position angle of line of nodes from nonlinear fits, (6) systemic velocity from nonlinear fits, 
(7) observed velocity dispersion, (8) velocity dispersion corrected for rotation, (9) ratio of rotation
velocity to velocity dispersion for the GCs, (10) power-law exponent of GC volume density profile, 
(11) mass from TME, (12) mass from rotation, (13) B-band mass to light ratio. Values in parentheses
give values calculated removing GC250 in VCC1261 and GC2 in VCC1087.}
\tablenotetext{a}{This value is slightly higher than the one quoted in Beasley et al. (2006) since 
we re-calculated the velocity dispersion here without attempting to correct for velocity uncertainties.}
\tablenotetext{b}{These values differ from those in Beasley et al. (2006) due to the above correction to 
the velocity dispersion.}
\end{deluxetable}

\end{document}